\documentclass[showpacs,preprint,aps,preprintnumbers,amsmath,amssymb]{revtex4}

\usepackage{epsfig}
\usepackage{graphicx}
\usepackage{dcolumn}
\usepackage{bm}

\let\jnfont=\rm
\def\NPB#1,{{\jnfont Nucl.\ Phys.\ B }{\bf #1},}
\def\PLB#1,{{\jnfont Phys.\ Lett.\ B }{\bf #1},}
\def\EPJC#1,{{\jnfont Eur.\ Phys.\ Jour.\ C }{\bf #1},}
\def\PRD#1,{{\jnfont Phys.\ Rev.\ D }{\bf #1},}
\def\PRL#1,{{\jnfont Phys.\ Rev.\ Lett.\ }{\bf #1},}
\def\MPLA#1,{{\jnfont Mod.\ Phys.\ Lett.\ A }{\bf #1},}
\def\JPG#1,{{\jnfont J.\ Phys.\ G}{\bf #1},}
\def\CTP#1,{{\jnfont Commun.\ Theor.\ Phys.\ }{\bf #1},}
\def\JHEP#1,{{\jnfont JHEP \ }{\bf #1},}
\def\NPPS#1,{{\jnfont Nucl.\ Phys.\ Proc.\ Suppl.\ }{\bf #1},}

\def\lsim{\mathrel{\mathpalette\oversim<}}
\def\gsim{\mathrel{\mathpalette\oversim>}}
\def\oversim#1#2{\lower0.5ex\vbox{\baselineskip0pt\lineskip0pt
  \lineskiplimit0pt\everycr{}\tabskip0pt
  \halign{$\mathsurround0pt #1\hfil##\hfil$\crcr #2\crcr\sim\crcr}}}
\begin{document}

\preprint{\parbox{1.2in}{\noindent arXive:0810.0989}}

\title{\ \\[10mm]
                  Current experimental constraints on NMSSM with large $\lambda$}

\author{\ \\[2mm]  Junjie Cao$^1$,  Jin Min Yang$^2$ }

\affiliation{$^1$ Ottawa-Carleton Institute for Physics, Department
                  of Physics, Carleton University, Ottawa, Canada K1S 5B6 \\
$^2$ Institute of Theoretical Physics and Kavli Institute for Theoretical Physics China,
Academia Sinica, Beijing 100190, China
     \vspace*{1.5cm}}

\begin{abstract}
The next-to-minimal supersymmetric model (NMSSM) with a large
$\lambda$ (the mixing parameter between the singlet and doublet
Higgs fields) is well motivated since it can significantly push up
the upper bound on the SM-like Higgs boson mass to solve the
little hierarchy problem. In this work we examine the current
experimental constraints on the NMSSM with a large $\lambda$,
which include the direct search for Higgs boson and sparticles at
colliders, the indirect constraints from precision electroweak
measurements, the cosmic dark matter relic density, the muon
anomalous magnetic moment, as well as the stability of the Higgs
potential. We find that, with the increase of $\lambda$,
parameters like $\tan\beta $, $M_A$,  $\mu$ and $M_2$ are becoming
more stringently constrained. It turns out that the maximal reach
of $\lambda$ is limited by the muon anomalous magnetic moment, and
for smuon masses of 200 GeV (500 GeV) the parameter space with
$\lambda \gsim 1.5 (0.6)$ is excluded.

\vspace*{1cm}
\end{abstract}


\maketitle

\section{introduction}
Since the minimal supersymmetric standard model (MSSM)
\cite{Haber} suffers from the $\mu$-problem \cite{muproblem} and
the little hierarchy problem, some non-minimal supersymmetric
models have recently attracted much attention, among which the
most intensively studied is the next-to-minimal supersymmetric
standard model (NMSSM) \cite{NMSSM}. In the NMSSM there is no
dimensionful parameters in the supersymmetry-conserving sector and
the $\mu$ term is dynamically generated through the coupling
between the two Higgs doublets and a newly introduced singlet
Higgs field which develops a vacuum expectation value of the order
of the SUSY breaking scale. The NMSSM provides two ways to
alleviate the little hierarchy problem. One is to relax the LEP II
lower bound on the mass of the SM-like Higgs boson, $h$, by
diluting  $ZZh$ coupling through the singlet component of $h$
and/or by suppressing the visible decay $h \to b\bar b $ through
introducing new decay of $h$ \cite{Gunion}. The other is to push
up the Higgs boson mass with a large $\lambda$, which can be
seen from the tree level upper bound of the Higgs boson mass
\cite{NMSSM-bound}
\begin{eqnarray}
m_{h, max}^2 \simeq m_Z^2 \cos^2 2 \beta + \lambda^2 v^2 \sin^2 2
\beta \label{limit}
\end{eqnarray}
where $\tan \beta = \langle H_u \rangle/\langle H_u \rangle $,
$v^2 = \langle H_u \rangle^2 + \langle H_u \rangle^2$ and
$\lambda$ is the mixing parameter between the singlet and doublet
Higgs fields defined in Eq.(\ref{Superpotential}).

Note that the choice of a large $\lambda$ to solve the little
hierarchy may be limited by the perturbativity of the theory at the
scale $\Lambda$ since the value of $\lambda$ is increasing with the
energy scale\cite{zerwas}. If this scale $\Lambda$ is the grand
unification (GUT) scale, $\lambda$ should be less than about 0.7 at
weak scale, leading to an upper bound on the Higgs boson mass of
about 150 GeV \cite{NMSSM-bound}. However, the bound on $\lambda$
from the perturbativity consideration can be relaxed by embedding
the NMSSM in some more complex frameworks. For example, in the Fat
Higgs model \cite{FatHiggs}, by completing the NMSSM (or NMSSM-like
models) with an appropriate strong dynamics at an intermediate scale
(much lower than the GUT scale), $\lambda$ can be as large as 2 at
weak scale and the Higgs boson mass can be pushed up to about 400
GeV. In this work, regardless the detailed forms of the ultraviolet
physics, we treat the NMSSM as an effective theory and examine the
current experimental constraints on its parameter space.

Such phenomenological studies on the Higgs boson and supersymmetry
are pressing since the mystery of the Higgs sector will
be unveiled at the LHC in the near future.
If the SM-like Higgs boson is
discovered with a mass above the MSSM upper bound, the NMSSM
(or other NMSSM-like models)  with a large $\lambda$, generally
called $\lambda$SUSY \cite{lamSUSY},  will be immediately favored
since it not only inherits all the advantages of the MSSM, such as
unifying gauge couplings and providing a dark matter candidate,
but also is free from the $\mu$-problem and the little hierarchy
problem. For the phenomenological studies of these models, a primary
work is to examine the current experimental constraints on their
parameter space.

We note that various constraints on the NMSSM have been studied in
the literature, but different constraints were considered in
different papers. For example, in \cite{NMSSM-Scan} the authors
mainly considered the LEP II constraints and put emphasize on
small $\lambda$ case. The package NMSSMtools \cite{NMSSMTools}
encoded various constraints (like the LEP II searches for the
Higgs boson, the cosmic dark matter and the stability of the Higgs
potential), but it is still not complete since it does not include
the indirect constraints from precision electroweak measurements
and the muon anomalous magnetic moment. In this work we consider
all these constraints and especially focus on the case with a
large $\lambda$. As will be shown from our study,  with the
increase of $\lambda$, the parameter space is getting more
stringently constrained. To figure out the allowed parameter space
is helpful for exploring such low energy supersymmetry at the LHC
and also may shed some light on constructing the ultraviolet physics
from the bottom-up view.

This paper is organized as follows. In Sec.II we briefly describe
the structure of the NMSSM with emphasis on its difference from the
MSSM. In Sec.III we summarize the constraints considered in
this work and briefly discuss their characters. In Sec. IV we scan
over the NMSSM parameter space and display the region allowed by
all these constraints. In Sec. V we give our conclusions.

\section{About the NMSSM}

The NMSSM extends the matter fields of the MSSM by adding one gauge
singlet superfield $\hat{S}$, and its superpotential takes the
form \cite{NMSSM}
\begin{eqnarray}
W & = & \lambda \varepsilon_{ij} \hat{H}_u^i \hat{H}_d^j \hat{S}  +
\frac{1}{3} \kappa  \hat{S}^3 + Y_u \varepsilon_{ij} \hat{Q}^i
\hat{U} \hat{H}_u^j - Y_d \varepsilon_{ij} \hat{Q}^i \hat{D}
\hat{H}_d^j - Y_e \varepsilon_{ij} \hat{L}^i \hat{E} \hat{H}_d^j
\label{Superpotential}
\end{eqnarray}
where $\hat{Q}$, $\hat{U}$ and $\hat{D}$ are squark superfields,
$\hat{L}$ and $\hat{E}$ are slepton superfields, $\hat{H}_u$ and
$\hat{H}_d$ are  Higgs doublet superfields.
The soft SUSY breaking terms are given by
\begin{eqnarray}
\label{vs} V_{\mbox{soft}} & = &
 \frac{1}{2} M_2 \lambda^a \lambda^a +\frac{1}{2} M_1 \lambda '\lambda '
 +m_d^2 |H_d|^2 + m_u^2 |H_u|^2+m_S^2 |S|^2  \nonumber \\
& & + m_Q^2 |\tilde{Q}|^2 + m_U^2 |\tilde{U}|^2 + m_D^2 |
\tilde{D}|^2
  +m_L^2 |\tilde{L}|^2 + m_E^2 |\tilde{E}|^2 \nonumber \\
& &  + (\lambda A_\lambda \varepsilon_{ij} H_u^i H_d^j S +
\mbox{h.c.})
     - (\frac{1}{3}  A_\kappa S^3 + \mbox{h.c.}) \nonumber \\
& & + (Y_u A_U \varepsilon_{ij} \tilde{Q}^i \tilde{U} H_u^j
   -Y_d A_D \varepsilon_{ij} \tilde{Q}^i \tilde{D} H_d^j
   -Y_e A_E \varepsilon_{ij} \tilde{L}^i \tilde{E} H_d^j
   +\mbox{h.c.}).  \label{soft}
\end{eqnarray}
Note that just like the MSSM, the NMSSM has the feature that SUSY
breaking induces the electroweak symmetry breaking. Before SUSY
breaking (i.e. without the soft breaking terms), the Higgs scalars
have zero vevs in the supersymmetric vacuum of the scalar
potential and thus the electroweak symmetry is not broken. After
SUSY breaking (i.e. with the soft breaking terms), the Higgs
scalars develop non-zero vevs in the physical (non-supersymmetric)
vacuum of the scalar potential and hence the electroweak symmetry
is spontaneously broken and the $\mu$ parameter is generated $\mu
= \lambda \langle S \rangle $. Since both the electroweak symmetry
breaking and the $\mu$ parameter generation are induced by SUSY
breaking, their scales should be naturally at the SUSY breaking
scale (the scale of soft breaking mass parameters).

The differences of the NMSSM and MSSM  come from the Higgs sector
and the neutralino sector\cite{NMSSM}. In the Higgs sector of the NMSSM
there are three CP-even and two CP-odd Higgs bosons.  In the basis
$[Re(H_u^0),Re(H_d^0), Re(S)]$, the mass-squared matrix elements for
CP-even Higgs bosons are
\begin{eqnarray}
{\cal M}_{S,11}^2 & = & m_A^2 \cos^2 \beta + m_Z^2 \sin^2 \beta,   \\
{\cal M}_{S,22}^2 & = & m_A^2 \sin^2 \beta + m_Z^2 \cos^2 \beta,   \\
{\cal M}_{S,33}^2 & = & \frac{\lambda^2 v^2}{4 \mu^2} m_A^2 \sin^2 2
\beta
  -\frac{\lambda \kappa}{2} v^2 \sin 2 \beta
  + \frac{1}{\lambda^2} \mu ( 4 \kappa^2  \mu - \lambda A_\kappa  ),  \\
{\cal M}_{S,12}^2 & = & (2 \lambda^2 v^2 - m_Z^2 - m_A^2 ) \sin \beta \cos \beta,   \\
{\cal M}_{S,13}^2 & = & 2 \lambda \mu v \sin \beta
  - \frac{\lambda v}{2 \mu} m_A^2 \sin 2 \beta \cos \beta - \kappa \mu v \cos \beta,  \\
{\cal M}_{S,23}^2 & = & 2 \lambda \mu v \cos \beta
   - \frac{\lambda v}{2 \mu} m_A^2 \sin \beta \sin 2 \beta - \kappa \mu v \sin \beta .
\label{CP-even}
\end{eqnarray}
In the basis $[\tilde{A}, Im(S)]$ with $\tilde{A} = \cos \beta~
Im(H_u^0) + \sin \beta~ Im(H_d^0)$, the mass-squared matrix elements
for the CP-odd Higgs bosons are
\begin{eqnarray}
\label{mA}
{\cal M}_{P,11}^2 & = & \frac{2
\mu}{\sin 2 \beta } \frac{\lambda A_\lambda +\kappa \mu}{\lambda} \equiv m_A^2, \\
{\cal M}_{P,22}^2 & = & \frac{3}{2} \lambda \kappa v^2 \sin 2 \beta
+ \frac{\lambda^2 v^2}{4 \mu^2} m_A^2 \sin^2 2 \beta  + \frac{3}{\lambda} \mu A_\kappa,  \\
{\cal M}_{P,12}^2 & = & \frac{\lambda v}{2 \mu} m_A^2 \sin 2 \beta
  - 3 \kappa \mu v.   \label{CP-odd}
\end{eqnarray}
As shown in Eq.(\ref{mA}), we can choose $m_A$ instead of  $A_\lambda$ as a free parameter.
So compared with the MSSM, the NMSSM has three additional parameters: $\lambda$, $\kappa$ and
$A_\kappa$. Conventionally, $\lambda$ is chosen to be positive
while $\kappa$ and $A_\kappa$ can be either positive or negative.
Note that Eqs.(\ref{CP-even}) and (\ref{CP-odd}) indicate that the parameters
$\lambda $ and $\kappa  \mu $ affect the mixings between doublet and singlet Higgs fields,
while $A_\kappa$ only affects the squared-mass of the singlet Higgs field.

In the neutralino sector, the NMSSM predicts one extra neutralino.
In the basis $(-i\lambda_1, - i \lambda_2, \psi_u^0, \psi_d^0, \psi_s )$ the  neutralino
mass matrix is given by \cite{NMSSM}
\begin{eqnarray}
\left( \begin{array}{ccccc}
M_1 & 0 & m_Z \sin \theta_W \sin \beta & - m_Z \sin \theta_W \cos \beta  & 0 \\
& M_2 & -m_Z \cos \theta_W \sin \beta & m_Z \cos \theta_W \cos \beta  & 0 \\
& & 0 & -\mu & -\lambda v \cos \beta \\
& & & 0 & - \lambda v \sin \beta \\
& & & & 2 \frac{\kappa}{\lambda} \mu \end{array} \right) .
\label{mass matrix}
\end{eqnarray}
This mass matrix is independent of $A_\kappa$, and the role of
$\lambda$ is to introduce the mixings of $\psi_s $ with  $\psi_u^0 $
and $\psi_d^0$, and $k \mu $ is to affect the mass of $\psi_s$. From
Eq.(\ref{CP-even},\ref{CP-odd},\ref{mass matrix}) one can learn
that in the limit $\lambda, \kappa \to 0 $, the singlet field have
no mixing with the doublet field and thus is decoupled.  In this
case, the NMSSM can recover the MSSM.

\section{constraints on the NMSSM parameters  }

Before we proceed to discuss experimental constraints on the
parameters of the NMSSM, we take a look at the bounds on $\lambda$
and $\kappa$ from the requirement that the theory should keep
perturbative under a certain scale $\Lambda$. The renormalization
group equations (RGEs) for $\lambda$ and $\kappa$ under the scale
$\Lambda$ take the following form \cite{RGE}
\begin{eqnarray}
  \frac{d \lambda}{d\ln\mu}
    &=& \frac{\lambda}{16 \pi^2} \left( 4 \lambda^2 + 2 \kappa^2
      + 3 Y_t^2 + 3 Y_b^2 + Y_\tau^2 - 3 g^2 - g'^2 \right),
\label{eq:RGElam} \\
  \frac{d \kappa}{d\ln\mu}
    &=& \frac{6\kappa}{16 \pi^2} \left( \lambda^2 + \kappa^2 \right),
\label{eq:RGEk}
\end{eqnarray}
where $g$ and $g'$ are the $SU(2)_L$ and $U(1)_Y$ gauge couplings.
These RGEs indicate that the values of $\lambda$ and $\kappa$
increase with the energy scale. The requirement of perturbativity
till the  cut-off scale $\Lambda$, i.e., $\lambda ( \Lambda ) \lsim
2 \pi$ and $\kappa ( \Lambda ) \lsim  2 \pi$, will set upper bounds
on $\lambda$ and $\kappa$ at weak scale (throughout this paper,
without specification all input parameters are defined at weak
scale). For example, if we assume that new dynamics appears at
$\Lambda = 10$TeV, we get $\lambda^2 + \kappa^2 \lsim 4.2 $ and for
$\lambda > 1.5$, $\kappa$ must be less than 1.2; while if $\Lambda$
is chosen to be the GUT scale, a stringent bound $\lambda^2+\kappa^2
\lsim 0.5$ is obtained \cite{zerwas}. In our following numerical
study we let $\lambda$ and $\kappa$ to vary below 2 and 1,
respectively, and this corresponds to set $\Lambda \simeq 10$ TeV.

In our study we consider the following constraints on the parameters of the NMSSM:
\begin{itemize}
\item[(1)] Constraints on the neutralino and chargino sector, which
include: the bound from invisible $Z$ decay
$\Gamma(Z \to \chi^0_1 \chi^0_1) < 1.76$ MeV;
the upper bounds on neutralino pair productions at LEP II
$\sigma(e^+e^- \to \chi^0_1 \chi^0_i) < 10^{-2}~{\rm pb}$ ($i>1$)
and $\sigma(e^+e^- \to \chi^0_i \chi^0_j) < 10^{-1}~{\rm pb}$; and
the LEP II bound on the lightest chargino mass $m_{\chi^+_1} > 103.5$ GeV.
These bounds will mainly constrain the parameters $M_{1}$, $M_2$ and $\mu$.

\item[(2)] Lower bounds on  sparticle masses from
           LEP and Tevatron experiments \cite{Yao}
\begin{eqnarray*}
&& m_{\tilde{e}} > 73  {\rm ~GeV},  \quad m_{\tilde{\mu}} > 94  {\rm
~GeV},
   \quad m_{\tilde{\tau}} > 81.9 {\rm  GeV}, \quad m_{\tilde{q}} > 250  {\rm ~GeV}, \\
&& m_{\tilde{t}} > 89  {\rm ~GeV}, \quad m_{\tilde{b}} > 95.7  {\rm
~GeV},
  \quad m_{\tilde{g}} > 195  {\rm ~GeV},
\end{eqnarray*}
where $m_{\tilde{q}}$ denotes the masses for the first two
generation squarks. These constraints will put lower bounds on
the soft breaking masses for sleptons and squarks.

\item[(3)] The LEP II lower bound on the charged Higgs boson mass,
           $m_{H^+} > 78.6$ GeV, which gives a lower bound on
           $m_A$ through the relation
           $m_{H^+}^2 = m_A^2 + M_W^2 - \frac{1}{2} \lambda^2 v^2$.

\item[(4)] Constraints from the direct search for Higgs boson at LEP II
\cite{Higgs}, which include various channels of Higgs boson productions
\cite{NMSSMTools}.  They will constrain the parameters $m_A$, $\tan \beta$,
$\lambda$ as well as the masses and the chiral mixing of top squarks
in a complex way.

\item[(5)] Constraint from the relic density of cosmic dark matter,
i.e. $ 0.0945 < \Omega h^2 < 0.1287 $ \cite{dmconstr}, assuming
the lightest neutralino is the dark matter particle.
 The relic density will constrain the parameters $M_1$, $M_2$,
$\mu$, $m_A$, $\tan \beta$ and $\lambda$ in a complex way \cite{darkmatter}.

\item[(6)] Constraint from the stability of the Higgs potential,
which requires that the physical vacuum of the Higgs potential with
non-vanishing vevs of Higgs scalars should be lower than any local
minima. Also, the scale of the Higgs soft breaking parameters should
not be much higher than the electroweak scale to avoid the
fine-tuning problem. Here we set 1 TeV as the upper bound of the
soft breaking parameters in the Higgs sector.  This will constrain
the parameters $m_A$,  $\mu$, $A_\kappa$, $\lambda$ and $\tan
\beta$.

\item[(7)] Constraints from precision electroweak observables such
as $\rho_{lept}$, $\sin^2 \theta_{eff}^{lept}$ and $M_W$, or their
combinations $\epsilon_i (i=1,2,3)$ \cite{Altarelli}. We require
the predicted $\epsilon_i$ in the NMSSM to be compatible with the
LEP/SLD data at $95.6\%$ confidence level or equivalently
$\chi^2/dof \leq 8.1/3 $.  We take the correlation coefficient of
$\epsilon_i$ from \cite{LEP} in calculating $\chi^2$. This
requirement will constrain the parameters $\tan\beta$, $m_A$ as
well as the soft breaking parameters in the third generation
squark sector.

\item[(8)] Constraint from $R_b = \Gamma (Z \to \bar{b} b) / \Gamma
( Z \to hadrons )$, whose measured value is $R_b^{exp} =
0.21629 \pm 0.00066 $ and the SM prediction is $R_b^{SM} = 0.21578 $
for $m_t = 173$ GeV \cite{Yao}. In our analysis we require
$R_b^{SUSY}$ is within the $2 \sigma$ range of its experimental
value. It has been shown that the SUSY contribution to $R_b$ might
be sizeable for large $\tan \beta $ \cite{Cao}.

\item[(9)] Constraint from the muon anomalous magnetic momentum $a_\mu$.
Now both the theoretical prediction and the experimental measurement
of $a_\mu$ have reached a remarkable precision, but they show a
significant deviation $a_\mu^{exp} - a_\mu^{SM} = ( 29.5 \pm 8.8 )
\times 10^{-10} $ \cite{Miller}. In our analysis we require the SUSY
effects to account for such deviation at $2 \sigma$ level. The
character of the SUSY contribution to $a_\mu$ is that it is
suppressed by smuon masses but enhanced by $\tan \beta$.
\end{itemize}

Among the above constraints, (1-6) and (9) have been encoded in the
package NMSSMTools \cite{NMSSMTools}.  In our calculations we extend
it by including the constraints (7) and (8).

The analytic expressions of $\epsilon_i$ and $R_b$ in
the NMSSM were given in our recent work \cite{Cao}. In  \cite{Cao} 
we also calculated the NMSSM contribution to $a_\mu$
(when we started that work, the results in \cite{Domingo,Gunion-g-2} 
had not yet published), where we 
extended the neutralino- and chargino-mediated MSSM contributions 
\cite{mu g-2} to the NMSSM  and also considered the contributions from the
Higgs-mediated diagrams \cite{mu-Higgs} and from the Barr-Zee diagrams
\cite{mu-Barr-Zee}. We checked that our $a_\mu$ results in \cite{Cao} 
agree with those in \cite{Domingo}.

Note that in our analysis we did not include the constraints from various
$B$-decays \cite{NMSSM-B} because they are dependent on squark flavor mixings
and thus involve additional parameters.

\section{Allowed regions of the NMSSM parameters}

In this section, we scan over the NMSSM parameter space to look for
the region allowed by the constraints in the preceding section.
Since we are interested in the parameters sensitive to the
constraints, we make some assumptions (as conservative as possible)
for the other parameters such as soft breaking parameters in squark,
slepton and gaugino sectors.

For the parameters in squark sector, we assume the so-called
$m_h^{max}$ scenario, which can maximize the lightest Higgs boson
mass \cite{mhmax}. This scenario assumes all the soft
breaking masses in the squark sector to be degenerate
\begin{eqnarray}
M_{\tilde{q}} = M_{Q_i} = M_{U_i} = M_{D_i}
\end{eqnarray}
with $i$ being the generation index. It also assumes  the trilinear
couplings to be degenerate $A_{u_i} = A_{d_i} $ with $( A_{u_i} -
\mu \cot \beta )/M_{\tilde{q}} = 2 $.  We fix $M_{\tilde{q}}= 1 TeV$
in our analysis since large $M_{\tilde{q}}$ can not only enhance the
lightest Higgs boson mass, but also decrease the contribution of the
third generation squarks to the electroweak parameters, which has the
same sign with the Higgs contributions \cite{lamSUSY}.
For the  parameters in slepton sector,
we note that the slepton masses affect little on the constraints
except the muon anomalous magnetic momentum.
In our calculation we assume all the soft breaking
parameters in the slepton sector are degenerate and take a
value of $200$ GeV (we will discuss the effects of its variation).
For the gaugino mass parameters, we
assume the grand unification relation
$M_1 = \frac{5}{3} ( g^{\prime 2}/g^2 ) M_2 $.

With the above assumptions, the free parameters are reduced to seven
($\lambda$, $\kappa$, $A_\kappa$, $\tan \beta$, $m_A$, $\mu$, $M_2$)
and within the capability of our computer to perform a scan.
During our scan, we first divide the varying range of $\lambda$ into
bins with each bin width being 0.1 and then we vary the values of
other parameters in the following ranges
\begin{eqnarray}
&& - 1 \leq \kappa \leq 1,  \quad \quad  1 \leq  \tan \beta \leq 60, \nonumber \\
&&   - 1 {\rm~TeV} \leq A_\kappa < 1 {\rm~TeV},
   \quad \quad 50 {\rm~GeV} \leq M_A, \mu, M_2 \leq 1 {\rm~TeV}.
\label{scan-region}
\end{eqnarray}

With two hundred million samples in each bin and keeping the points
satisfying the constraints, we finally get the
allowed regions of these parameters. Our scan results indicate that
the number of the survived samples for $\lambda < 0.5$
is much larger than that for $\lambda > 0.5$, which means
that the parameters for small $\lambda$ are much less constrained
than the case with large $\lambda$.
Since we are interested in large $\lambda$, here we only
show our scan results for $\lambda > 0.5$.

\begin{figure}[tbp]
\epsfig{file=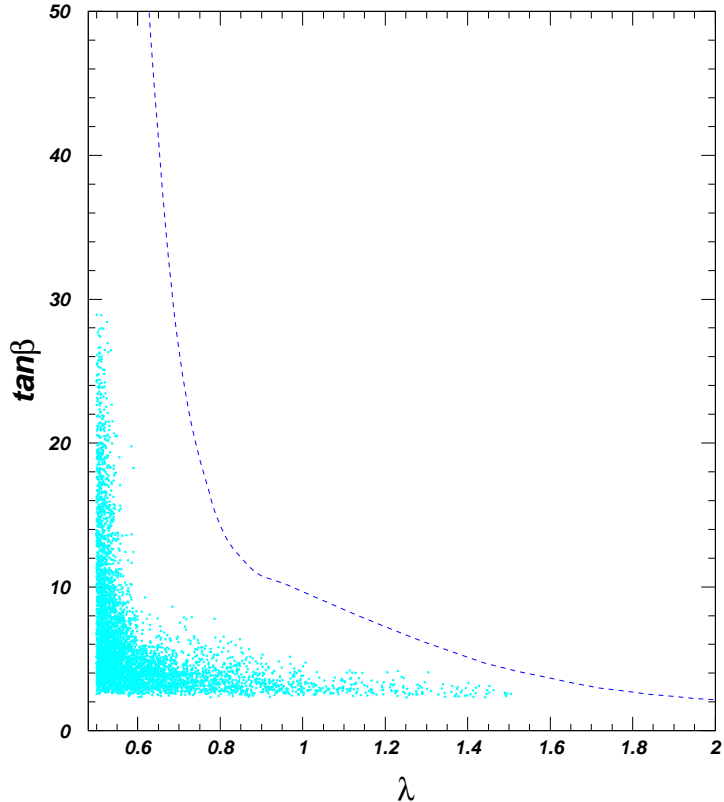,width=10cm} \vspace{-0.7cm} \caption{\small
The scatter plots are the NMSSM parameters satisfying all the
constraints (1-9). The curve is the upper bound on $\tan \beta$
without considering the muon g-2 constraints.} \label{lam-tanb}
\end{figure}

In Fig.\ref{lam-tanb} we display the parameters (scatter plots)
satisfying all the constraints (1-9) in the plane of
$\lambda$ versus $\tan \beta$.  Also, we present a curve which is
the upper bound on $\tan \beta$ without considering the muon g-2 constraints.
To get this curve, we fix $\lambda$ and scan over the parameters in
Eq.(\ref{scan-region}). We adopt the important sampling
method \cite{Lepage} to optimize the varying range of $\tan \beta$.

Fig.\ref{lam-tanb} shows that the upper bound on
$\tan \beta$ gets stronger  as $\lambda$ gets large,
and when all the constraints are considered,
$\lambda$ is upper bounded by about 1.5.
The underlying reason for this
is that the constraints (1-8), especially the
constraint (7), have limited the maximal value of $\tan \beta $,
which decreases with the increase of $\lambda$.
Since a large $\tan \beta$ is needed to explain the deviation
of the muon g-2, $\lambda$ must terminate at a certain value
where the corresponding $\tan \beta$ value is too small to explain
the muon g-2.
We have checked that the maximal value of $\lambda$ is dependent on slepton mass.
For example, for slepton mass of $100$ GeV,  $280$ GeV and  $500$ GeV,
the bounds on $\lambda$ are $\lambda\lsim 2$,  $\lambda\lsim 1$ and
$\lambda\lsim 0.6$, respectively.

\begin{figure}[tbp]
\epsfig{file=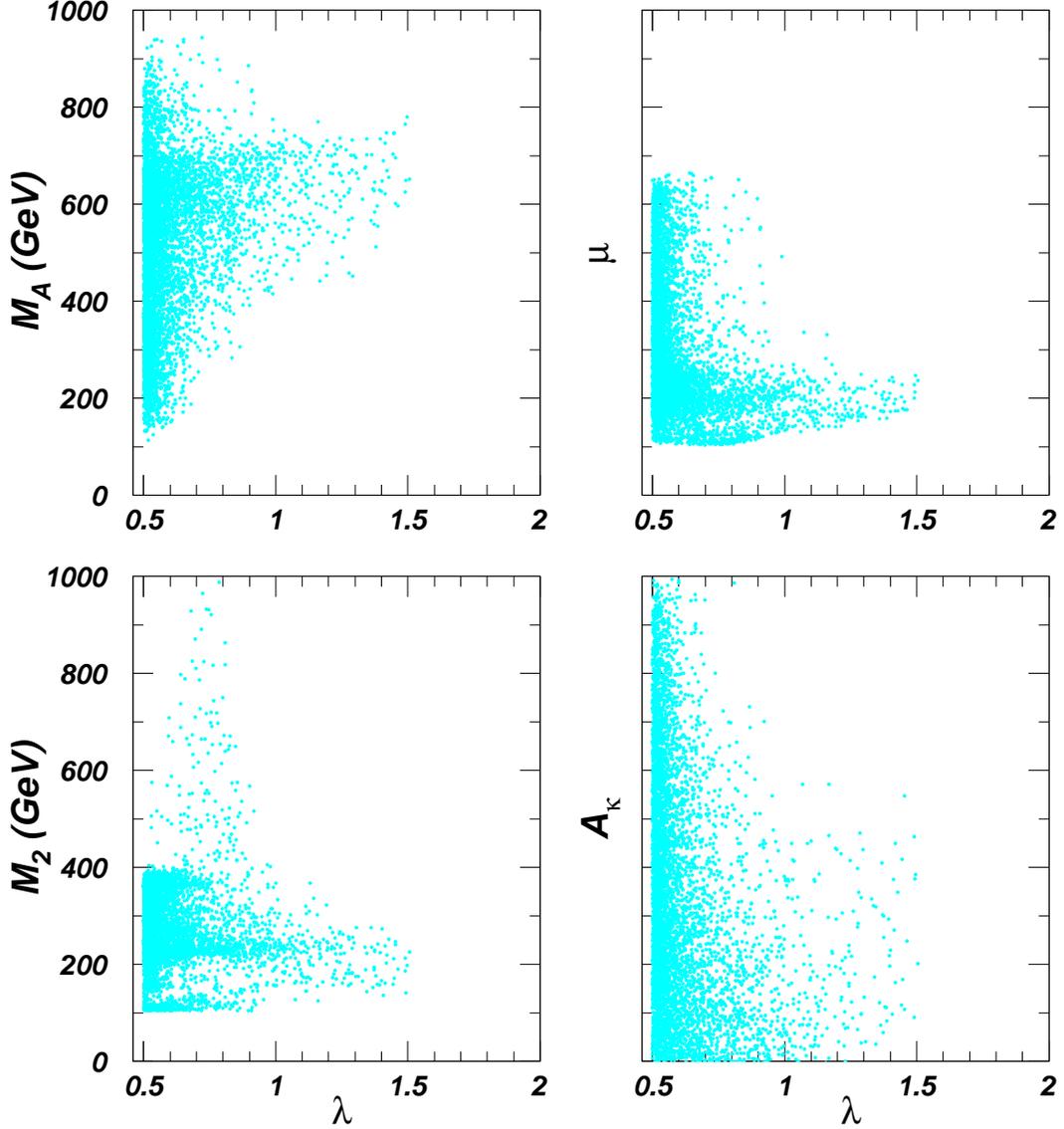,width=15cm} \vspace{-0.7cm} \caption{\small
Scatter plots of the NMSSM parameters satisfying all the
constraints (1-9), displayed in different planes.}
\label{lambda-all}
\end{figure}

In Fig.\ref{lambda-all} we display the NMSSM parameters satisfying
all the constraints in different planes. We see that for a large
$\lambda$ the parameters $m_A$, $\mu$, $M_2$ and $A_\kappa$ are
also bounded in a certain region. For $\lambda =1$, these bounded
regions are $400 {\rm ~GeV} \lsim  M_A \lsim  800$ GeV, $150 {\rm
~GeV} \lsim  \mu \lsim  250$ GeV, $150 {\rm ~GeV} \lsim M_2 \lsim
300$ GeV  and $A_\kappa \lsim  600$ GeV.

From the figure of $M_A$ versus $\lambda$ in Fig.\ref{lambda-all}
one can see that the lower bound of $M_A$ increases as $\lambda$
becomes large. The reason is that the LEP II direct search for Higgs
boson mainly limits the mass and the couplings of the light CP-even
Higgs boson whose component is dominated by the doublet Higgs field
$H_u$ or $H_d$. For $\tan\beta > 1$, this Higgs boson should be
dominantly composed by $H_u$ field since ${\cal M}_{S,11}^2 $ is
smaller than ${\cal M}_{S,22}^2$, and its mass is to be reduced by
the off-diagonal elements ${\cal M}_{S,12}^2 $ and ${\cal
M}_{S,13}^2$. As $\lambda$ gets larger, these off-diagonal elements
get larger and hence reduce the mass of the light CP-even Higgs
boson, which then requires a larger $M_A$ to compensate in order to
satisfy the LEP II lower bound.

The figure of $\mu$ versus $\lambda$ in Fig.\ref{lambda-all}
indicates that with the increase of $\lambda$, the upper bound of
$\mu$ decreases. This is because in the off-diagonal elements
${\cal M}_{S,13}^2 $ and ${\cal M}_{S,23}^2$ (which reduce the
light CP-even Higgs boson mass), $\lambda $ is always
associated with $\mu$,  and to meet the LEP II bound a large $\lambda$
must be accompanied by a small $\mu$.

The figure of  $M_2$ versus $\lambda$ in Fig.\ref{lambda-all}
shows that $M_2$ is also bounded in a narrow region.
This is because the relic density of the dark matter
correlates the parameters $m_A$, $\mu$, $M_2$, $\lambda$ and $\tan
\beta$ in a complex way, and a large value for any of these
parameters will limit severely the region of other
parameters.

The figure of  $A_\kappa$ versus $\lambda$ in Fig.\ref{lambda-all}
shows that the trilinear soft breaking parameter  $A_\kappa$
for the singlet field is also limited. This can
be understood from the expressions of ${\cal M}_{S,33}^2 $
and ${\cal M}_{P,22}^2 $.
The stability of the Higgs potential requires both of them to be
positive, which sets an double-sided bound on $A_\kappa$.

We also studied the relationship between the Yukawa couplings
$\lambda$ and $\kappa$, and we found no correlation between them.
Even for $\lambda = 1.5$, the value of $\kappa$  can still vary from
0.3 to 1.

Next, we take a look at the Higgs boson masses allowed by the
constraints. Since a large $\lambda$ can enhance the lightest
CP-even Higgs boson mass and thus avoid the little hierarchy
problem, it is interesting to look at the dependence of the Higgs
boson masses on the parameter $\lambda$.

\begin{figure}[tbp]
\epsfig{file=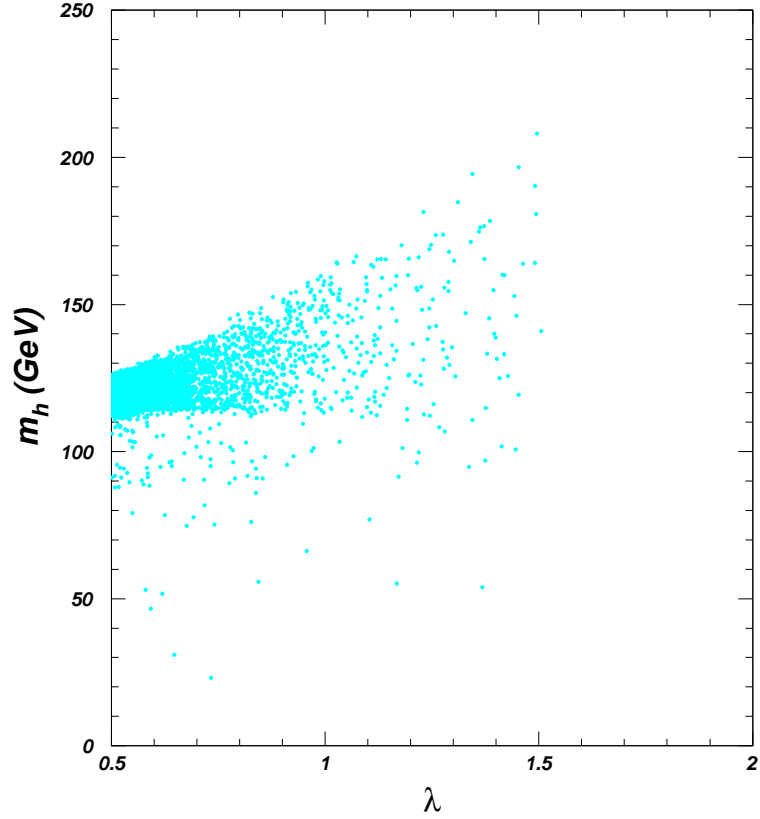,width=10cm} \vspace{-0.7cm} \caption{\small
Same as Fig.\ref{lambda-all}, but for $\lambda$ versus
         the lightest CP-even Higgs boson mass $m_h$.}
\label{lambda-mh}
\end{figure}

\begin{figure}[tbp]
\epsfig{file=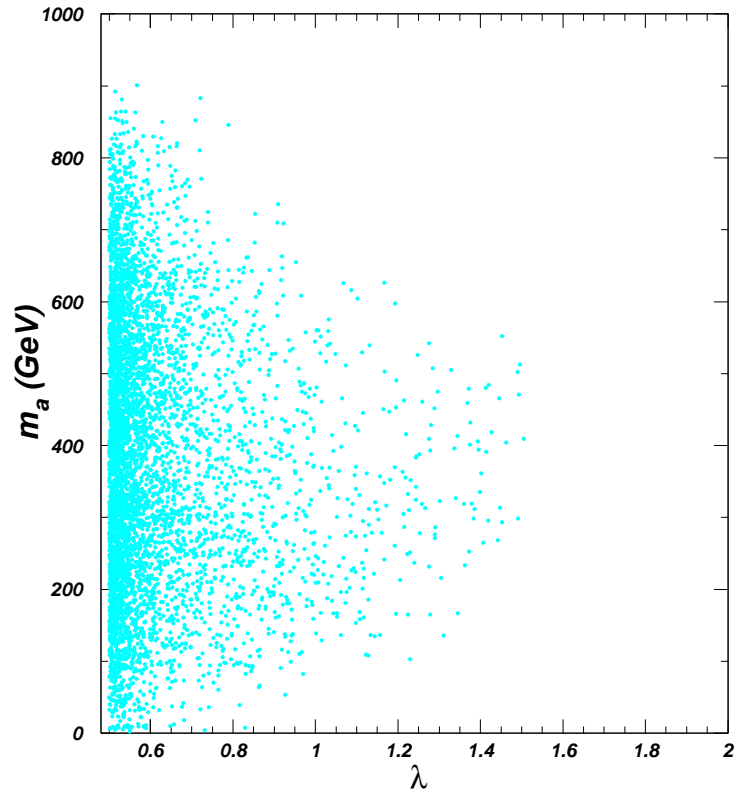,width=10cm} \vspace{-0.7cm} \caption{\small
Same as Fig.\ref{lambda-all}, but for $\lambda$ versus
         the lightest CP-odd Higgs boson mass $m_a$.}
\label{lambda-ma}
\end{figure}

In Figs.\ref{lambda-mh} and \ref{lambda-ma} we show our scan
results in $\lambda$ versus $m_h$ plane and $\lambda$ versus $m_a$
plane with $m_h$ being the lightest CP-even Higgs boson mass and
$m_a$ the lighter CP-odd Higgs boson mass. From
Fig.\ref{lambda-mh} one can learn that the upper bound of $m_h$
increases with $\lambda$, which is expected from Eq.(\ref{limit}),
and for $\lambda = 1.5$ the value of $m_h$ can reach 210 GeV. From
Fig.\ref{lambda-ma} one can learn that with the increase of
$\lambda$, a super light CP-odd Higgs boson is gradually ruled
out, and for $\lambda > 1 $ it is bounded in the range $100
{\rm~GeV} \lsim m_a \lsim 600$ GeV. The properties of these Higgs
bosons can be quite different from those in the MSSM, and their
phenomenology at the LHC was discussed in \cite{Cavicchia}.

\begin{figure}[tbp]
\epsfig{file=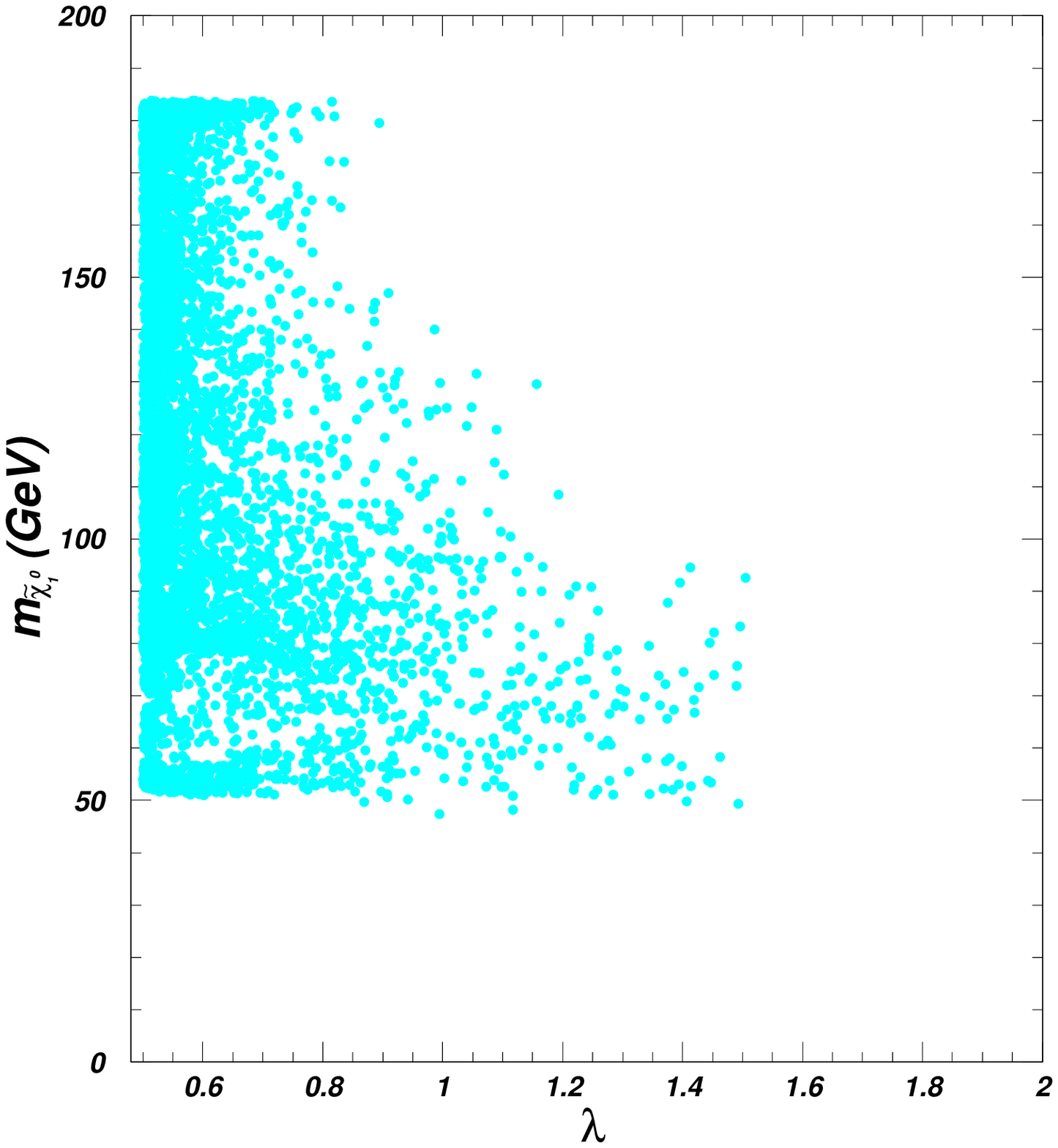,width=10cm} 
\vspace{-0.7cm} 
\caption{\small Same as Fig.\ref{lambda-all}, but for
         the lightest neutralino mass $m_{\tilde{\chi}^0_1}$
         versus $\lambda$.}
\label{lambda-chi}
\end{figure}
\begin{figure}[tbp]
\epsfig{file=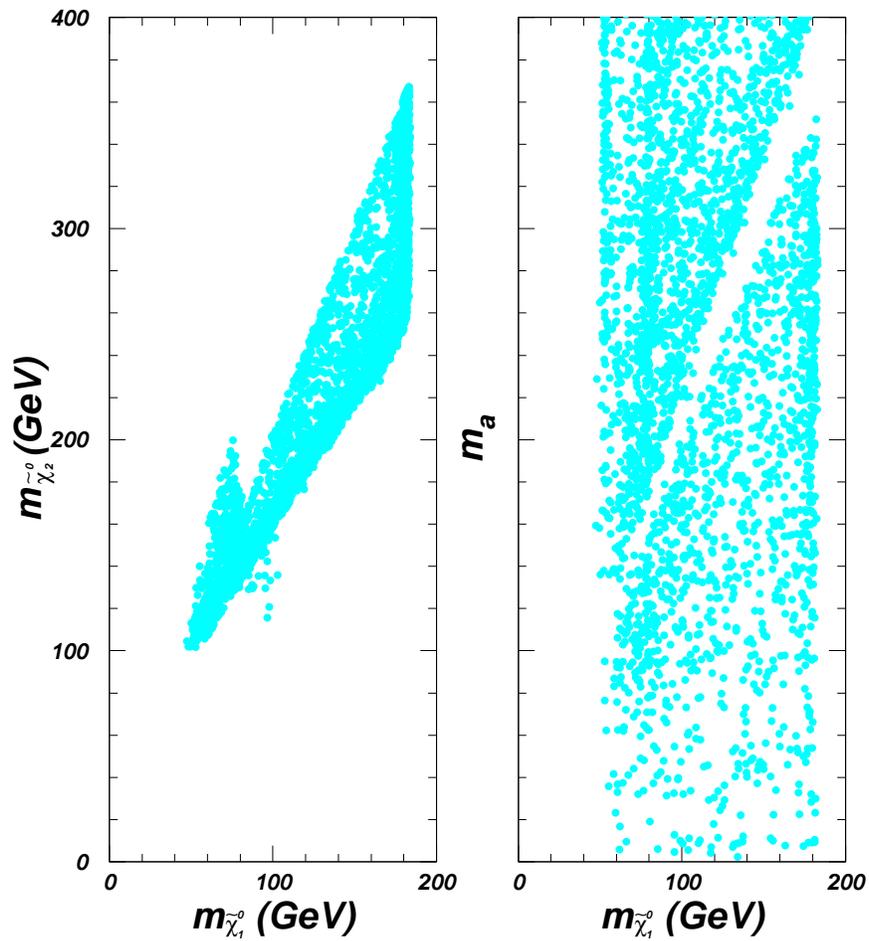,width=12cm} 
\vspace{-0.7cm} 
\caption{\small  Same as Fig.\ref{lambda-all}, but for
 $m_{\tilde{\chi}^0_2}$ and $m_a$ versus $m_{\tilde{\chi}^0_1}$. } 
\label{spectrum}
\end{figure}

Finally, in order to understand the mechanism used to reproduce the correct 
dark matter abundance, we consider the properties in the neutralino sector. 
In the NMSSM with large $\tan \beta$, the component of the lightest 
neutralino is either
higgsino-dominant or bino-dominant for a light 
mass below 80 GeV, but for a heavier mass it is bino-dominant.
In Fig.\ref{lambda-chi} we show our scan results in the plane
of $m_{\tilde{\chi}^0_1}$ versus $\lambda$. We see that 
with the increase of $\lambda$, the upper bound on 
$m_{\tilde{\chi}^0_1}$ becomes stringent and eventually
it is constrained in the range of $50 \sim 100$ GeV. 
About the next lightest neutralino $\tilde{\chi}^0_2$ 
we found that its mass is constrained in the range of 
$100 \sim 160$ GeV for $\lambda > 1.2$.  
In order to figure out the annihilation mechanism of 
$\tilde{\chi}^0_1$ in providing for the dark matter relic density,
we compare the masses of $\tilde{\chi}^0_2$ and $a$ with
$\tilde{\chi}^0_1$ in Fig.\ref{spectrum}. 
This figure indicates that $\tilde{\chi}^0_2$
is significantly heavier than $\tilde{\chi}^0_1$.
Since in our scan the slepton masses are fixed to 200 GeV,
also significantly heavier than $\tilde{\chi}^0_1$, 
we conclude that the coannihilation of $\tilde{\chi}^0_1$ 
with $\tilde{\chi}^0_2$ or with a slepton is generally 
Boltzmann-suppressed and plays an unimportant role
in accounting for the dark matter relic density. 
Note that, as shown in Fig.\ref{spectrum}, there are some
samples around the funnel region $2 m_{\tilde{\chi}^0_1} \sim
m_a$ and in this case the annihilation of $\tilde{\chi}^0_1$ 
through the s-channel exchange of a light 
$a$ becomes dominant \cite{Gunion-darkmatter}.

\section{conclusion}
The NMSSM with a large $\lambda$ is an attractive scenario since
it can push up the upper bound on the SM-like Higgs boson mass to
solve the little hierarchy problem. We examined the current
experimental constraints on this scenario, which include the
direct experimental bounds, the indirect constraints from
precision electroweak measurements, the cosmic dark matter relic
density, the muon anomalous magnetic moment, as well as the
stability of the Higgs potential. Our results showed that for a
large $\lambda$ the parameter space is severely  constrained. For
example, for a smuon mass of 200 (500) GeV the parameter space
with $\lambda \gsim 1.5 (0.6)$ is excluded, and for $\lambda=1$
the allowed ranges are $2.5 \sim 4$ for  $\tan\beta$,
$400 \sim 800$ GeV  for $M_A$, $150 \sim 250$ GeV for $\mu$, 
$150 \sim 300$ GeV for $M_2$ 
and $0 \sim 600$  GeV for  $A_\kappa$.

Finally, we would like to point out that our conclusion may be
qualitatively applicable to other NMSSM-like models such as the
Minimal Nonminimal Supersymmetric Standard Model (MNMSSM)
\cite{n-MSSM}, which has similar structure with the NMSSM and can
be viewed as the low energy realization of the Fat Higgs model
\cite{FatHiggs}. For example, it has been pointed out that for any
singlet extensions of the MSSM, regardless the form of its
superpotential, a large $\lambda$ is always accompanied by a small
$\tan \beta$ \cite{lamSUSY}.  This property, as shown in our
paper, can either limit the smuon mass or limit $\lambda$ if we
require the theory to explain the deviation of the muon anomalous
magnetic momentum. Another example is about the constraint from
dark matter. In the MNMSSM we expect that the constraint can limit
the relevant parameters in a more stringent way than in the NMSSM
since the neutralino sector in the MNMSSM is exactly same as in
the NMSSM but with fixed $\kappa =0$ \cite{Menon}.

\section*{Acknowledgement}
This work was supported in part by the National Sciences and
Engineering Research Council of Canada, and by the National
Natural Science Foundation of China (NNSFC) under grant No.
10505007, 10725526 and 10635030.


\end{document}